\documentstyle[emulateapj]{article}
\newcommand{\zdot}{\makebox[0pt][l]{.}}
\newcommand{\up}[1]{\ifmmode^{\rm #1}\else$^{\rm #1}$\fi}

\newcommand{\uph}{\up{h}}
\newcommand{\upm}{\up{m}}
\newcommand{\ups}{\up{s}}
\newcommand{\arcd}{\ifmmode^{\circ}\else$^{\circ}$\fi}
\newcommand{\arcm}{\ifmmode{'}\else$'$\fi}
\newcommand{\arcs}{\ifmmode{''}\else$''$\fi}

\begin{document}

\def\thefootnote{\fnsymbol{footnote}}

\title{Photometric detection of high proper motions
in dense stellar fields using Difference Image Analysis}

\author{L. Eyer \& P.R. Wo\'zniak}

\affil{Princeton University Observatory, Princeton, NJ 08544, USA}
\affil{leyer@astro.princeton.edu, wozniak@astro.princeton.edu}

\begin{abstract}

The Difference Image Analysis (DIA) of the images obtained by the Optical
Gravitational Lensing Experiment (OGLE-II) revealed a peculiar artifact
in the sample of stars proposed as variable by Wo\'zniak (2000) in one of
the Galactic bulge fields: the occurrence of pairs of candidate variables
showing anti-correlated light curves monotonic over a period of 3 years.
This effect can be understood, quantified and related to the stellar proper
motions. DIA photometry supplemented with a simple model offers an effective
and easy way to detect high proper motion stars (HPM stars) in very dense
stellar fields, where conventional astrometric searches are extremely
inefficient.

\end{abstract}


\section{Introduction}

\label{sec:intro}

The Optical Gravitational Lensing Experiment (OGLE) is a long term massive
photometric search for microlensing events (Udalski et al. 1992). Due to
intrinsically low probability of this phenomenon microlensing searches
focus on monitoring the densest stellar fields of the sky, the Galactic
bulge region and the brightest galaxies of the Local Group (e.g. Paczy\'nski
1996a). In 1997 the OGLE project entered its second phase (OGLE-II) when
the new, dedicated 1.3m Warsaw Telescope at the Las Campanas
Observatory\footnotemark, Chile, started operation
(Udalski, Kubiak \& Szyma\'nski 1997).
The temporal and spatial coverage were significantly expanded, currently
reaching a total of 11 square degrees in 49 bulge fields, and also including
almost entire bar regions of LMC and SMC. The data is gradually being released
on the Internet.

\footnotetext{The Las Campanas Observatory is operated by the Carnegie
Institution of Washington}

Wo\'zniak (2000) started Difference Image Analysis (DIA) of the OGLE-II images.
Using a new technique invented by Alard \& Lupton (1998) and further developed
by Alard (2000) it is possible to match seeings and intensity scales of two
images of the same stellar field and obtain a meaningful difference image.
For variable objects in crowded environments the photometry on difference
frames is superior to conventional PSF fitting.
%
%
Three seasons of data for the first OGLE-II bulge field,
BUL\_SC1, are now in public domain (Wo\'zniak 2000).

The field BUL\_SC1 contains about 850,000 stars detectable with DoPhot
software (Schechter, Mateo \& Saha 1995) on the $2048\times 8192$ pixel
reference image (a stack of 20 best frames). It covers $0.24 \times 0.95$
deg$^2$ and is centered at $\alpha_{2000}=18\uph02\upm32\zdot\ups5$,
$\delta_{2000}=-29\arcd57\arcm41\arcs$. Corresponding galactic coordinates
are $l=1\zdot\arcd08$, $b=-3\zdot\arcd62$. The median FWHM seeing is
$1.3\arcs$ in the full set of 197 images obtained during 3 observing
seasons between 1997 and 1999. The frames are taken in drift-scan mode using
a $2048\times 2048$ pixel SITe3 chip with the approximate pixel scale of
$0.417\arcs{\rm pixel^{-1}}$ (Udalski, Kubiak \& Szyma\'nski 1997).

The observed stars have apparent $I$ magnitudes between 11.5 and 20.5
with rapidly decreasing completeness below $I=19.5$. They can be highly
blended. The list of candidate variable stars in that field contains 4597
objects. The DIA analysis of these data revealed a peculiar effect, which
was first interpreted as a potential problem, and later turned out to be
a discovery. Some candidate variables seemed to come in pairs separated
roughly by 1 FWHM of the seeing disk. Upon visual inspection of selected cases,
we noticed that light curves of stars in such couples showed approximately
linear and anti-correlated trends over the entire 3 year period. This kind
of long term correlated behavior would be very unlikely an instrumental
artifact. Because the centroid of each candidate variable is determined once
for the entire series of images, we hypothesized that a star having
a detectable proper motion may produce two
"variable stars" in difference frames, even if the star has a constant
magnitude during the observed period. After we confirmed this interpretation,
we learned that Drake et al. (2001) have been working on the same problem
with the MACHO database of conventional PSF photometry, although they
proceeded in the opposite direction. Following the theoretical suggestion
of Andrew Drake they found that some constant stars had the same type of
temporal trend in observations: parabolic fall-off with increasing scatter,
as predicted for stars with detectable motions. Both searches are compared
in Section~\ref{sec:fits}. Similar possibilities must be
hiding in the OGLE database of standard PSF photometry.

In this paper we discuss this effect and present a list of selected stars
with detectable proper motions. We propose a method for selecting high
proper motion stars (hereafter HPM stars) in dense stellar fields using
difference imaging and briefly discuss possible applications.

\section{The signature of high proper motions \\ in the DIA photometry}

\label{sec:signature}

\begin{figure}[t]
\plotfiddle{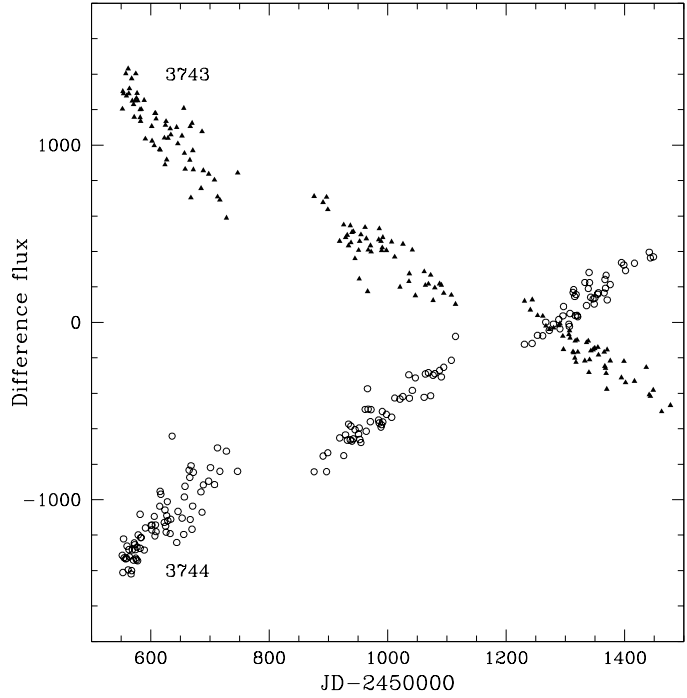}{8.5cm}{0}{50}{50}{-150}{-80}
\caption{\label{fig:lc}
   The DIA light curves of two candidate variables in the OGLE-II bulge field
   BUL\_SC1. The difference flux between each frame and the reference frame
   is shown as a function of the truncated Julian day. The stars are separated
   by $1.2\arcs$ and are not physical sources. They are an artifact produced by
   a single moving object. Note that the flux differences for both members
   of the pair cross zero near day 1249, which is the mean epoch of the frames
   used to construct the reference frame. 
}
\end{figure}

As we mentioned before among candidate variables detected using DIA
pipeline by Wo\'zniak (2000) there is a noticeable number of close pairs
separated in the frame by the distance comparable to the FWHM of the PSF.
Light curves of both members of the pair are anti-correlated and approximately
linear in time for 3 years covered by observations. In Figure~\ref{fig:lc}
we show a clear example of this effect.


\subsection{Simple model}

\label{sec:model}

In the DIA method, a reference image is subtracted from each individual
image in a series of observations for a given field. This can be done after
all images are resampled to the common pixel grid. Coordinate system is fixed
with respect to the positions of stars in the dominant stellar population
in one of the images called the coordinate template, in this case the red
clump stars at $I\sim15.5$ mag. Therefore for imaging towards the Galactic
bar, the reference coordinate system is strongly weighted towards stars
at about 8.5 kpc away. Proper motion of the Galactic Center at
$\sim6$ mas yr$^{-1}$ is common to all stars in the Galactic bar region.
The majority of stars in the image should have totally negligible relative motions.
The presence of the foreground disk stars with magnitudes around $I=15.5$
complicates slightly this definition, but the main assumption certainly holds
to the level needed in this preliminary analysis. A better defined reference
system can be used in the future, however it involves major changes in the software. 

\begin{figure}[t]
\plotfiddle{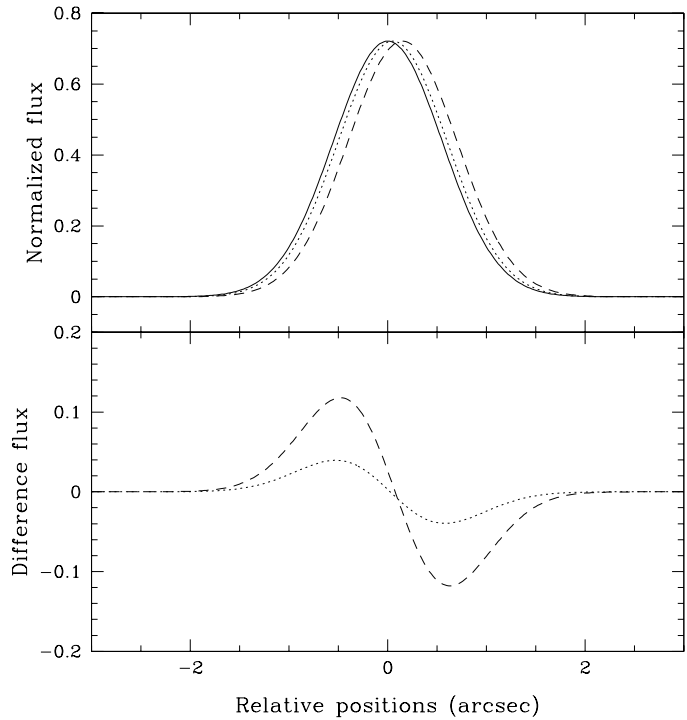}{8.5cm}{0}{50}{50}{-150}{-80}
\caption{\label{fig:gauss}
Schematic view of the image subtraction process for a moving
star using a Gaussian PSF in one dimensional space. In the top panel
we show the PSF in the reference image (solid line) and in two images taken
at later epochs (dotted and dashed lines respectively). Below there are
corresponding residual patterns. For small displacements the amplitude
of the residual increases linearly with separation.
}
\end{figure}

\begin{figure}[t]
\plotfiddle{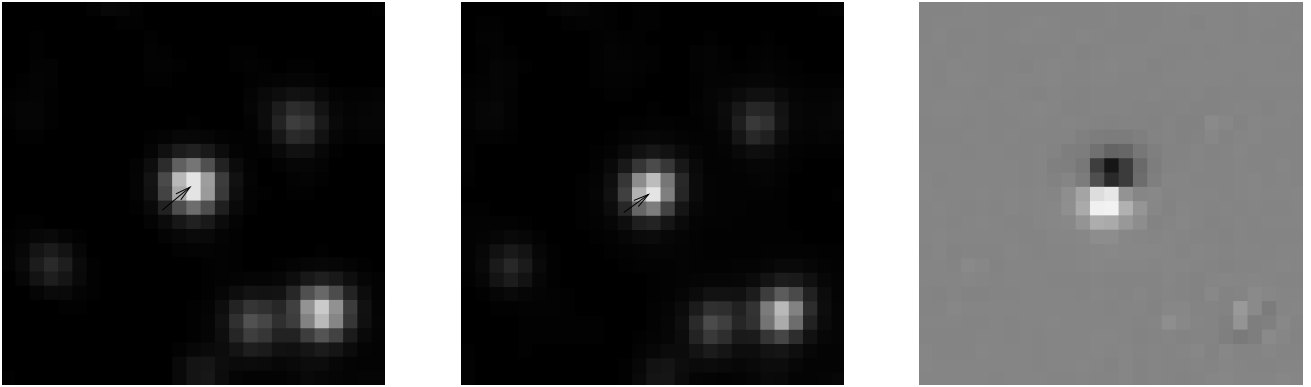}{8.5cm}{-90}{80}{80}{-320}{350}
\caption{\label{fig:hpm_field}
The signature of a high proper motion star in difference imaging.
We show the averages of 4 early frames (left), 4 late frames (middle),
and their difference (right). Between early and late epochs separated
by 920 days the star near the center moved by about 0.5 pixels, as indicated
with arrows. Note that the moving star cannot be missed in the difference frame,
while finding the same motion in regular frames requires careful examination.
The scale is $0.417\arcs$pixel$^{-1}$ and the star moves at
$\mu_{t}=83$ mas yr$^{-1}$.
}
\end{figure}

Let us suppose that one of the stars in the field (most likely a nearby one)
has a detectable proper motion. Its centroid will be generally shifted with
respect to the reference image. When the reference image is then subtracted
from a given program image a wavy pattern will appear near the position
of the moving star. Figure~\ref{fig:gauss} illustrates this phenomenon
in one dimensional space using a Gaussian approximation to the PSF. In the
top panel we schematically show the PSF of the moving star in three different
images taken at three different epochs. The difference flux after subtraction
of the reference PSF from the two remaining images is shown in the bottom
panel. Clearly the amplitude of the residual increases as the star moves
away from its template position. The residual left in difference frames by
stars with proper motions in a certain range has a characteristic dipole
shape, approximately antisymmetric with respect to the middle point
between the extrema of the two, positive and negative, components. For moderate
proper motions the separation of the poles in this dipole is larger than the
positional shift of the star between epochs in question and both poles have
characteristic width of the PSF, but their shape is not strictly that of the PSF.
When the shift is already comparable to, or larger than the FWHM, the residual
simply consists of two separate PSF components. This issue will be discussed
further in the context of sensitivities (Section~\ref{sec:fits}). In all cases
the axis of the dipole is aligned with the apparent direction of the star
velocity vector in the sky. An example of this type of residual is shown
in Figure~\ref{fig:hpm_field} for a star with proper motion of 83 mas yr$^{-1}$.

The residual in the form similar to the PSF and detectable in several consecutive
frames will be included in the list of candidate variables by the DIA pipeline
(Wo\'zniak 2000). It is clear from Figure~\ref{fig:hpm_field} that numerous cases
of HPM stars can be found this way because both components of the dipole
have a good chance to make it as suspected variables.


The total flux in the dipole will be zero within the errors as long as
the HPM star does not vary itself, and therefore at any given time the sum
of the differences flux in both components of the pair will also stay near zero.
The zero flux difference for an individual light curve will occur at
the mean time of the template frames. In case of the BUL\_SC1 field most
template frames were taken during the third season, which perfectly agrees with
the flux difference crossing zero around JD=2451249 (Figure~\ref{fig:lc}).


The trend in flux can be related to the proper motion $\mu$ in a straightforward
way. The photometric detection of two objects in the difference frame for a pair
of two individual frames at times separated by $t$ can be approximated as coming
from the subtraction of two Gaussians:

\begin{equation}
\label{eq:model}
   f(x) = \frac{F_{\mbox{\tiny tot}}}{\sqrt{2 \pi} \sigma}
          exp^{-\frac{(x-\mu t)^2}{2 \sigma^2}}
        - \frac{F_{\mbox{\tiny tot}}}{\sqrt{2 \pi} \sigma}
          exp^{-\frac{x^2}{2 \sigma^2}},
\end{equation}

\noindent
where $F_{\mbox{\tiny tot}}$ is the total flux, and $\mu t$ is centroid the shift.
In order to imitate photometric measurements of Wo\'zniak (2000) we expand
the right side of Equation~\ref{eq:model} for small $\mu t/\sigma$, take the time
derivative and integrate over spatial dimension between $x=0$ to $x=\infty$
to obtain the slope of the light curve $\gamma={\rm d} F(t)/{\rm d} t$. We find
that the proper motion $\mu$ in this linear approximation is related to $\gamma$ by:

\begin{equation}
\label{eq:estimate}
\mu = \frac{\sqrt{2 \pi} \sigma}{F_{\mbox{\tiny tot}}} \gamma .
\end{equation}

The result remains valid in a two dimensional case. For each HPM star we can
choose the coordinate system such that the displacement and the dipole axis
of symmetry will be along the $x$ axis. The equation in two dimensions
integrated between $y=-\infty$ and $y=\infty$ is then given, again, by
Equation~\ref{eq:model}.

\subsection{Parallax and differential refraction}

\label{sec:parref}

In Figure~\ref{fig:lc} it can be seen that the individual slopes in each
season are different from the general slope over all three seasons.
There are at least two additional effects which
change the relative position of the star with respect to the remaining stars in
the field. One of them, the parallactic motion, is of great interest because
it gives a distance estimate for relatively nearby stars, and HPM stars are
likely to be sufficiently close. The other is differential refraction, the
wavelength dependent centroid shift, which must be considered in fine astrometric
work (e.g. Binnendijk 1960).

An order of magnitude estimate of the parallax can be made for a typical thick
disk star with the transverse velocity $v_t=33$ km/s. We have
$v_t=4.77~\mu/\pi$, and then $\mu/\pi \approx 7$. This effect is therefore
rather small, nevertheless it could be seen in some of our cases and is worth
considering. In a general field a close star will revolve around the ellipse
in the reference frame of distant stars, for which the parallax is negligible.
The ellipse is determined by the ecliptic coordinates of the star and the
motion should obviously have a one year period. Interestingly, the Galactic bulge
(Sagittarius: $\alpha\sim 18\uph$, $\delta\sim -30\arcd$) lies almost
on the ecliptic at $\lambda=270\arcd$, $\beta=-6\arcd$ in the ecliptic system.
In this case the parallactic ellipse is strongly flattened and the ecliptic
longitude $l$ component dominates (Binnendijk 1960). Furthermore because
the ecliptic is nearly parallel to the celestial equator there and the CCD chip
is aligned with the axes of the equatorial system, the expected centroid shifts
in our data are much stronger in $x$ direction than in $y$ ($\alpha$ vs. $\delta$).
Also, the best time of the year to observe the bulge is near its opposition
with respect to the Sun, which means that the extrema of the parallactic
displacement will be poorly covered by observations, in particular
by observations under good seeing conditions and at low zenith angles $z$.

Differential refraction decreases the apparent zenith angle of the
observed object by the amount which depends on the color of the object.
Because of the wavelength dependence of the atmospheric refractive index,
the images of stars are in fact short spectra with the dispersion axis
directed towards the zenith, and therefore blue objects are affected more.
For general theory see e.g. Kovalevsky (1995).
Alcock et al. (1999) discussed the effect in the context of the DIA
photometry on MACHO images. The MACHO blue filter is sufficiently broad
that the refraction alone is capable of producing wavy residuals in
difference frames and substantially increasing the photometric scatter.
In case of the MACHO data a statistical correction was applied using color
maps of each field. However it should be stressed that for individual cases
with locally atypical colors this correction cannot work, and may even have
wrong sign. The only way around this problem is the use of narrow filters. 
In MACHO red band data the situation is dramatically better
(Alcock et al. 1999) and in our $I$ band data with the standard
Johnson-Kron-Cousins filter the effect is very small.
Nevertheless the quality of the centroid for $I$=11--13 mag stars in OGLE images
is sufficiently good to reveal the influence of the color on the apparent position,
and we include in the model the color dependent term linear in $\tan(z)$.
The projected shift in equatorial coordinates $\alpha$, $\delta$ changes
with the zenith angle and hour angle. Somewhat unexpectedly, an asymmetry
arises here. In relatively large part of the sky around the meridian the shift
in $\delta$ has null time derivative, while the shift in $\alpha$ is proportional
to hour angle and dominates. Also the sign of the shift in $\alpha$ depends on whether
we look east or west, which results in unfavorable correlation
with time of the year, despite the fact that the zenith angle is
uncorrelated with date in the observing strategy adopted by OGLE. This
explains a cumbersome coupling between refraction and parallax
(Section~\ref{sec:fits}).

\section{Results}

\label{sec:results}

\subsection{Selection criteria}

\label{sec:criteria}

\begin{figure}[t]
\plotfiddle{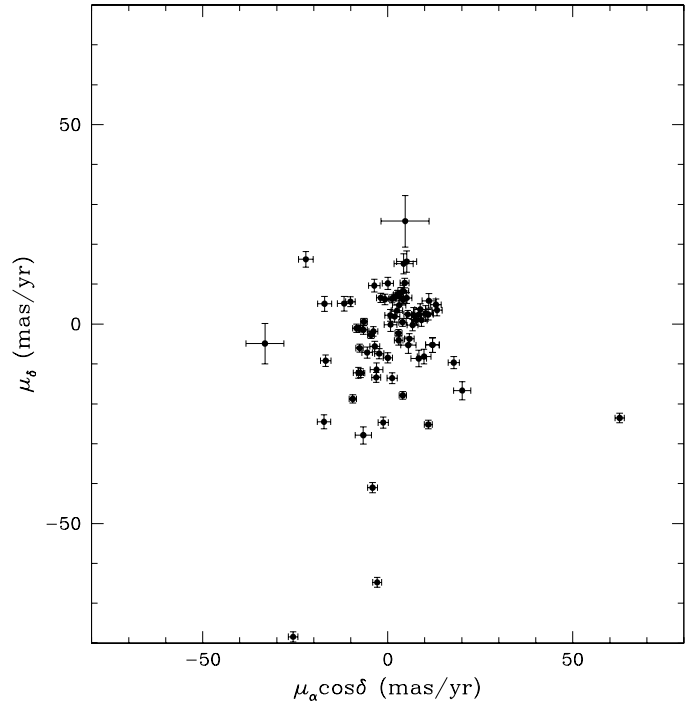}{8.5cm}{0}{50}{50}{-150}{-80}
\caption{\label{fig:dist2d}
Two dimensional distribution of proper motions $\mu_{\alpha}$,
$\mu_{\delta}$ of stars in Table~\ref{tab:stars}.
}
\end{figure}

We start by selecting all close couples in the database of candidate variables
of Wo\'zniak (2000). We only select couples in which the two components have
light curve slopes of the opposite sign. It is also required that the absolute
value of the ratio of the two slopes is between 0.1 and 10.0. The significance
of the slope is assessed using the Student test. There were 99 couples
separated by less than 4.5 pixels with less than 0.01 probability of insignificant
slope in any of the components. There is a small gap in the distribution of
separations around 4--5 pixels, and there are no interesting cases at higher
separations.  The next step is an attempt to detect the HPM star near the
middle position between both (spurious) stars of the pair. After measuring
positions with respect to the neighboring stars at all possible epochs, we
perform the Fisher test comparing models with and without the proper motion.
Parallax and refraction are not considered at this point. In the final sample
we admit 74 stars with less than 1 per thousand probability that the fit improvement
with the addition of proper motion is not significant. For these candidates
we fit the full model including parallactic motion and refraction.
Table~\ref{tab:stars} contains $I$ magnitudes and fitted proper motions of
stars which passed all criteria. In Figure~\ref{fig:dist2d} we show the
two dimensional distribution of fitted proper motions. Total proper motions
$\mu$ typically range from 4 mas yr$^{-1}$ to 60 mas yr$^{-1}$.

\subsection{Model fits and sensitivity}

\label{sec:fits}

\begin{figure}[t]
\plotfiddle{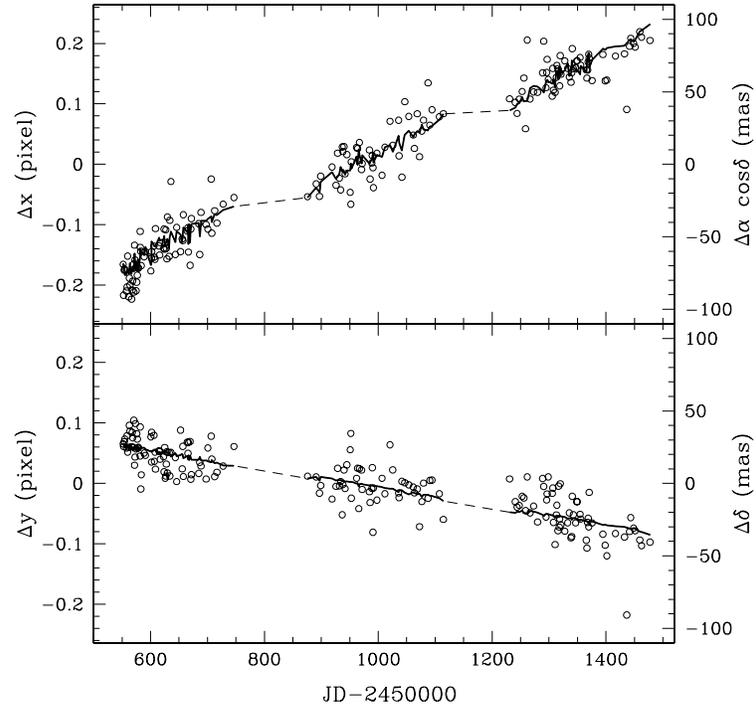}{8.5cm}{0}{50}{50}{-150}{-80}
\caption{\label{fig:example}
Model fit to the coordinates of the star 3743-3744, a photometrically
detected high proper motion star. Differential refraction accounts
for most of the residuals around the straight line model and its
influence is stronger in $x$.
}
\end{figure}

\begin{figure}[t]
\plotfiddle{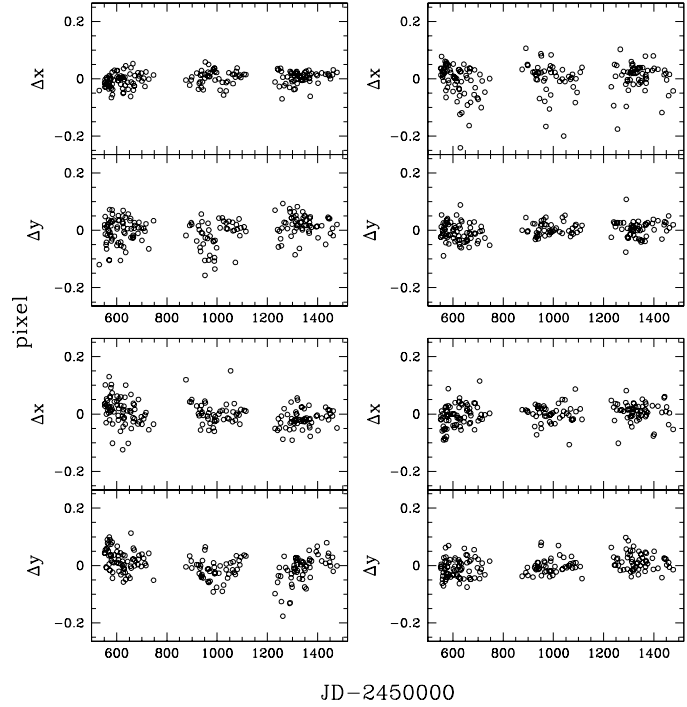}{8.5cm}{0}{50}{50}{-150}{-80}
\caption{\label{fig:comp}
Coordinates of four stars in the neighborhood of the star 3743-3744 in a 
3 year observing period. The positions of stars defining the frame
of reference are stable.
}
\end{figure}

\begin{figure}[t]
\plotfiddle{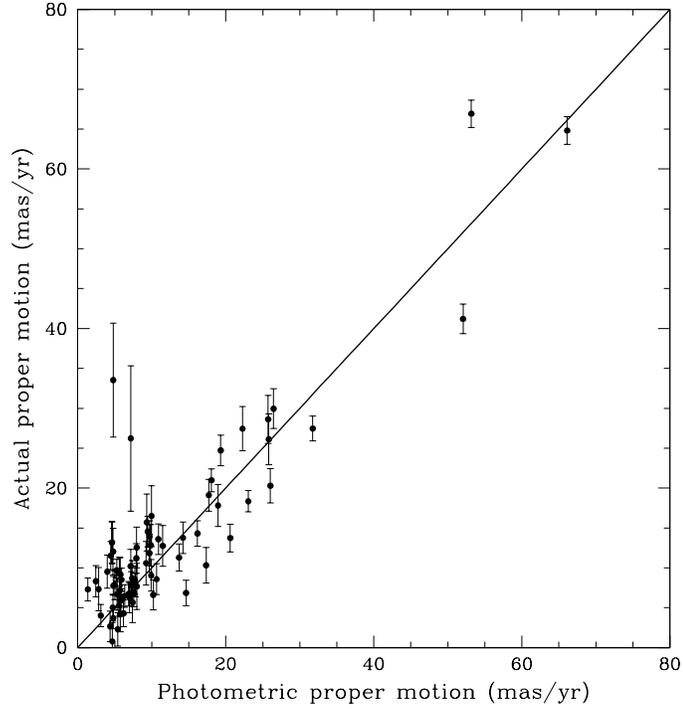}{8.5cm}{0}{50}{50}{-130}{-80}
\caption{\label{fig:model}
Predictions of Equation~\ref{eq:estimate} versus the actual measured
proper motion for HPM stars in Table~\ref{tab:stars}. Taking into account
that seeing variations increase the scatter in this plot, the DIA photometry and
the orientation of the dipole residual give a fairly accurate estimate
of $\mu_{\alpha}$ and $\mu_{\delta}$. Two outliers with large error bars
are both in very tight blends and have uncertain total fluxes.
}
\end{figure}

\begin{figure}[t]
\plotfiddle{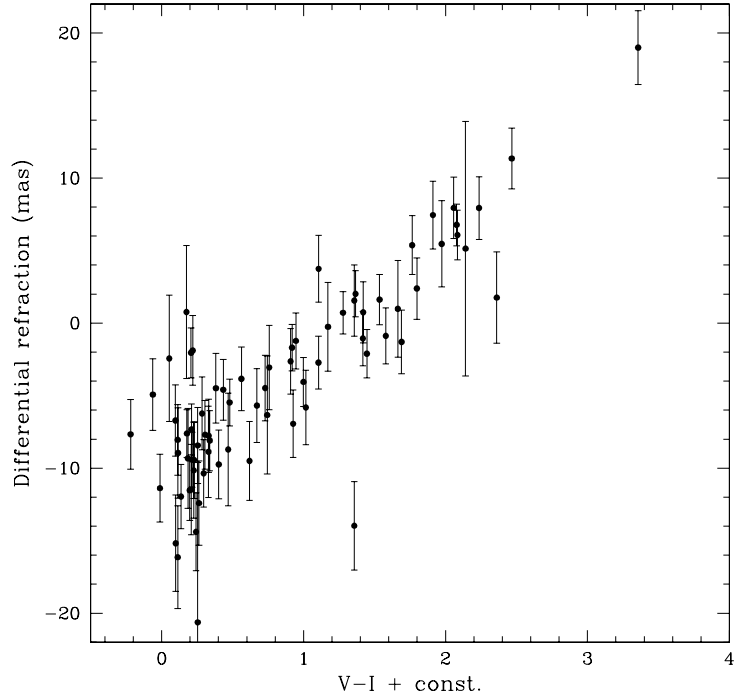}{8.5cm}{0}{50}{50}{-130}{-80}
\caption{\label{fig:color}
Correlation between the colors for stars in Table~\ref{tab:stars}
and the fitted shift due to refraction (relative vertical displacement
at zenith angle $45\arcd$). Similarly to Figure~\ref{fig:model},
outliers are members of tight blends.
}
\end{figure}

Figure~\ref{fig:example} presents the data and the model for a star with clearly
detected proper motion. We made sure that most stars do not show similar trends;
4 examples can be found in Figure~\ref{fig:comp}. For the purpose of this preliminary
announcement we neglect the fact that the mapping between the template frame coordinates
and the equatorial system may be locally warped. This hardly affects any of our
conclusions. Typical scatter around the
fit to the coordinate transformation between OGLE-II frames is about 0.06 pixels
($0.024\arcs$) in each coordinate (Wo\'zniak 2000). Stars in our HPM sample
are on average brighter than stars used to match pixel grids in the DIA
pipeline and have r.m.s. position uncertainty of about 0.03 pixels ($0.012\arcs$).
In principle we should be able to detect some stars with measurable parallaxes.
However, as discussed in Section~\ref{sec:parref}, the influence of
differential refraction is correlated with the date of observation
and results in the fitting degeneracy between the parallax and the refraction.
Additionally, we observe systematic residuals in our fits at 1$\sigma$ level,
suggesting changes in the state of the atmosphere in the observed direction
between the beginning and the end of the season. This could be due to either
seasonal effects or differing air conditions towards the land and towards
the ocean. Because refraction dominates the residuals after the effects of
proper motion are taken out, we cannot report convincing parallax values.
Nevertheless, the measured proper motion agrees quite well with the predictions
of Equation~\ref{eq:estimate}, as demonstrated in Figure~\ref{fig:model}.
Keeping in mind that the photometric estimate for $\mu$ in
Equation~\ref{eq:estimate} depends on seeing, (FWHM$=\root\of{8\ln 2}~\sigma$),
we get surprisingly good information about the direction and magnitude of $\mu$
from fixed centroids and light curves of both components of the dipole residual,
without a single actual positional measurement. Figure~\ref{fig:color} confirms
that the fitted amplitude of the refractive displacement is, indeed, well correlated
with the (uncallibrated) $V-I$ color of the star. This relation could be used
in the future for calculating rather than fitting the displacement due
to refraction.

In our search almost all discovered HPM stars have $I$ magnitudes
between 12 and 14, which suggests that the present method is not sensitive
to faint objects, although the full range of colors is covered.
Drake et al. (2001) report discoveries all the way to $V\sim19$ using
a different method based on regular PSF photometry, but they also mention
that in five years of data the lower limit on detectable proper motion was
about 30 mas yr$^{-1}$. They find 154 stars in the area of 50 square degrees.
We see numerous good detections below 30 mas yr$^{-1}$ in 3 years of observations
of $\sim0.2$ square deg, suggesting that the two approaches cover different
parts of the parameter space. It should be also pointed out that the reference
image composed of frames from a broad range of epochs is not optimal for this
kind of work. It tends to wash out the signal by stretching the reference PSF
of the moving star. Without modifications our method will not be capable of
discovering stars with very high proper motions, which quickly move 2$\times$FWHM
away from the location on the reference image and produce entirely
different "light curves".

\section{Discussion and future perspectives}

\label{sec:discussion}

The effects of the centroid shift in difference image photometry for BUL\_SC1
OGLE field appeared as a serendipity phenomenon. We propose to use this clear
signature to detect HPM stars in very crowded fields. In its present version
described here, the method is insensitive to very high proper motions,
however the potential for improvement is great. With the suitable selection
of difference images obtained from a series of observations covering
sufficiently long base line, it should be possible to achieve much better
efficiency and expand the range of detectable $\mu$. Perhaps the use of light
curves in the detection process can be eliminated altogether, and one should
concentrate on preparation of good combined reference frames around a number
of epochs along the whole observing sequence. Difference frames between epochs
separated by various time intervals will reveal the same type of dipole-like residual
near stars with broad range of velocities.

The OGLE survey, and other photometric surveys, offer thus promising astrometric
results. Methods like ours, or the one used by Drake et al. (2001) for the
MACHO data, make it possible to filter out HPM stars out of millions of
constant stars in crowded fields with very little extra effort. Crowding
practically prevented conventional astrometric surveys like NLTT (New Luyten
Two Tenths catalogue) from detecting stars in the Galactic bulge. It is generally
depleted around $b=0$. NLTT goes down to the photographic magnitude $m_R=19.2$,
but has a detection limit of about 180 mas yr$^{-1}$. In a small area we found
37 stars with proper motions between 10 and 80 mas yr$^{-1}$. Tentative
predictions for OGLE-II as a whole can be made. There are 4 observing seasons
and 49 Bulge fields. All these fields will be reduced by the DIA pipeline
(Wo\'zniak 2000), and we could therefore expect the determination of more than
1500 proper motions from the OGLE-II bulge data alone without any software
modifications. It is likely that after fine tuning the technique will return
significantly larger sample of HPM stars.

The possibility that large numbers of HPM stars in crowded fields may be
known within 1--2 years has interesting implications for some proposed
experiments to detect astrometric microlensing. The cross section for astrometric
displacement falls off as the impact parameter to $-1$ power, much slower than
power $-4$ in case of the photometric disturbance (Miralda-Escud\'e 1996 and
references therein). Paczy\'nski (1996b) showed that high proper motion stars
should within a reasonable time cause predictable microlensing events. In case
of closest moving lenses and much fainter than the source (contrast can be
increased with the proper filter), the astrometric shift could be measured
with the HST and would provide a direct determination of the lens mass. Some
of the local lenses could be white or brown dwarfs. Of course with the
1 $\mu$as accuracy of the Space Interferometry Mission (SIM) a lot more is
possible. The described experiment using SIM would provide up to 1\% accuracy
in the mass determination (Paczy\'nski 1998). Another possibility discussed
by Paczy\'nski (1998) is the determination of stellar radii (or equivalently
effective temperatures) from microlensing events which are astrometric
and have resolved sources due to very low impact parameter. According to SIM
specification, the degeneracy between physical microlensing parameters can be
broken for events out to the distance of the Galactic bulge and even LMC and
SMC, provided that they are both astrometric and photometric
(e.g. Paczy\'nski 1998). Gould (2000a) proposed to measure the mass function
of stellar remnants in the Galactic bulge. Samir \& Gould (2000) and Gould (2000b)
discuss the selection of candidate objects for SIM. The method proposed here would
greatly extend the list of possible targets.


\begin{acknowledgements}

We would like to thank Prof. Paczy\'nski for communicating enthusiasm
and useful comments. L.E. would like to thank Prof. Grenon, Dr Pfenniger
and Dr Pourbaix for valuable discussions and for the hospitality of the
Geneva Observatory, where this article was partially written. This work
was supported partly with the NSF grant AST--9820314 to Bohdan Paczy\'nski
and the grant from the Swiss National Science Foundation to L.E.

\end{acknowledgements}


\newpage

\newpage

\newpage

\newpage

\newpage

\newpage

\newpage

\newpage

\newpage


\begin{deluxetable}{ccccrrrrrr}
\tablecaption{\label{tab:stars} High proper motion stars found in the OGLE bulge field SC1.}
\tablewidth{18cm}
\tablehead{
\colhead{Star} &
\colhead{$I$} &
\colhead{$\alpha_{2000}$} &
\colhead{$\delta_{2000}$} &
\colhead{$\mu_{\alpha\ast}$\tablenotemark{\dagger}} &
\colhead{$\sigma_{\mu_{\alpha\ast}}$} &
\colhead{$\mu_{\delta}$} &
\colhead{$\sigma_{\mu_{\delta}}$} &
\colhead{$\mu_{t}$} &
\colhead{$\sigma_{\mu_{t}}$}
\vspace{2mm} \nl
\colhead{ID} &
\colhead{mag} &
\colhead{$\uph~~\upm~~\ups$} &
\colhead{$\arcd~~\arcm~~\arcs$} &
\multicolumn{6}{c}{mas yr$^{-1}$}
}
\startdata
  21 -   22 &  14.13 &  18:02:25.23 &  $-$30:25:09.1 &     7.0 &  2.0 &     2.2 &  1.9 &    7.3 &  2.7  \nl
  95 -  110 &  14.17 &  18:02:25.70 &  $-$30:24:25.1 &     5.3 &  2.7 &    15.0 &  2.6 &   15.9 &  3.8  \nl
 121 -  133 &  13.94 &  18:02:38.63 &  $-$30:24:40.3 &  $-$3.5 &  1.3 &  $-$5.6 &  1.3 &    6.6 &  1.9  \nl
 171 -  172 &  14.24 &  18:02:24.59 &  $-$30:23:22.4 &    10.2 &  1.9 &     2.7 &  1.9 &   10.6 &  2.7  \nl
 187 -  198 &  14.24 &  18:02:34.98 &  $-$30:23:49.6 &     4.3 &  2.5 &    15.1 &  2.5 &   15.7 &  3.6  \nl
 290 -  291 &  13.13 &  18:02:05.40 &  $-$30:22:11.9 &     5.1 &  1.4 &     6.6 &  1.4 &    8.3 &  1.9  \nl
 348 -  357 &  13.79 &  18:02:54.88 &  $-$30:22:00.1 &  $-$7.7 &  0.8 &  $-$1.2 &  0.8 &    7.8 &  1.2  \nl
 360 -  384 &  13.63 &  18:02:00.94 &  $-$30:21:12.6 &     5.4 &  1.9 &  $-$5.1 &  1.9 &    7.5 &  2.7  \nl
 371 -  372 &  13.39 &  18:02:11.32 &  $-$30:20:53.6 &     4.0 &  1.0 &     0.5 &  1.0 &    4.0 &  1.4  \nl
 376 -  377 &  13.46 &  18:02:12.99 &  $-$30:20:55.5 &  $-$0.4 &  1.5 &    10.2 &  1.5 &   10.2 &  2.1  \nl
 382 -  450 &  12.87 &  18:02:15.31 &  $-$30:20:37.7 &    17.8 &  1.5 &  $-$9.7 &  1.5 &   20.3 &  2.1  \nl
 458 -  459 &  15.08 &  18:02:19.86 &  $-$30:19:54.2 &  $-$2.9 &  1.2 & $-$64.8 &  1.2 &   64.8 &  1.7  \nl
 513 -  514 &  13.81 &  18:02:06.58 &  $-$30:19:21.3 &     8.4 &  1.6 &     2.4 &  1.6 &    8.7 &  2.2  \nl
 519 -  520 &  13.93 &  18:02:14.07 &  $-$30:19:39.9 &     9.1 &  1.5 &     1.0 &  1.5 &    9.2 &  2.1  \nl
 624 -  625 &  13.93 &  18:02:51.27 &  $-$30:18:28.1 &     4.8 &  6.5 &    25.8 &  6.4 &   26.2 &  9.1  \nl
 628 -  637 &  12.44 &  18:02:52.52 &  $-$30:18:24.0 &    11.3 &  1.8 &  $-$5.3 &  1.7 &   12.5 &  2.5  \nl
 630 -  631 &  11.67 &  18:02:53.84 &  $-$30:18:01.5 &  $-$9.4 &  1.0 & $-$18.7 &  1.0 &   21.0 &  1.4  \nl
 664 -  665 &  14.06 &  18:02:21.70 &  $-$30:17:18.1 &     0.8 &  1.7 &  $-$0.1 &  1.7 &    0.8 &  2.4  \nl
 670 -  677 &  12.88 &  18:02:40.40 &  $-$30:17:57.9 &     2.9 &  1.2 &  $-$4.5 &  1.2 &    5.3 &  1.7  \nl
 749 -  750 &  13.30 &  18:02:47.71 &  $-$30:16:42.1 &     2.5 &  1.6 &     3.4 &  1.5 &    4.2 &  2.2  \nl
 804 -  805 &  14.60 &  18:02:19.13 &  $-$30:15:25.0 &     1.2 &  1.4 & $-$13.6 &  1.3 &   13.6 &  1.9  \nl
 809 -  824 &  13.49 &  18:02:34.47 &  $-$30:15:36.9 & $-$33.2 &  5.1 &  $-$4.9 &  5.0 &   33.5 &  7.2  \nl
 820 -  821 &  13.91 &  18:02:44.46 &  $-$30:16:04.6 &     4.1 &  1.0 & $-$17.9 &  0.9 &   18.3 &  1.3  \nl
 840 -  841 &  13.01 &  18:03:02.38 &  $-$30:16:11.3 &  $-$0.8 &  1.4 &     6.2 &  1.3 &    6.2 &  1.9  \nl
 896 -  906 &  12.86 &  18:02:59.46 &  $-$30:14:58.0 &    10.9 &  1.3 &     2.4 &  1.3 &   11.2 &  1.9  \nl
 977 -  978 &  14.42 &  18:02:02.96 &  $-$30:13:31.3 &  $-$2.3 &  1.3 &  $-$7.4 &  1.3 &    7.7 &  1.8  \nl
1012 - 1013 &  12.91 &  18:02:35.04 &  $-$30:13:11.4 &     8.4 &  2.1 &  $-$8.6 &  2.0 &   12.0 &  2.9  \nl
1065 - 1071 &  13.92 &  18:02:12.81 &  $-$30:12:14.0 &  $-$9.5 &  1.2 &     5.2 &  1.2 &   10.9 &  1.7  \nl
1107 - 1108 &  14.03 &  18:02:45.14 &  $-$30:12:02.4 &  $-$3.0 &  1.7 & $-$11.4 &  1.7 &   11.8 &  2.4  \nl
1203 - 1204 &  13.71 &  18:02:58.57 &  $-$30:11:32.3 & $-$25.6 &  1.3 & $-$78.4 &  1.3 &   82.5 &  1.8  \nl
1351 - 1359 &  13.37 &  18:02:25.10 &  $-$30:08:30.7 &     2.2 &  1.0 &     6.8 &  1.0 &    7.2 &  1.4  \nl
1379 - 1380 &  13.55 &  18:02:55.46 &  $-$30:08:51.2 &  $-$4.1 &  1.3 & $-$41.0 &  1.3 &   41.2 &  1.9  \nl
1410 - 1411 &  14.51 &  18:02:18.69 &  $-$30:07:24.1 &     5.4 &  1.0 &     2.4 &  1.0 &    5.9 &  1.4  \nl
1538 - 1539 &  13.00 &  18:02:25.94 &  $-$30:06:01.5 &     4.1 &  1.0 &     5.8 &  1.0 &    7.1 &  1.4  \nl
1592 - 1593 &  14.19 &  18:02:08.53 &  $-$30:05:23.1 &     2.9 &  0.9 &  $-$2.3 &  0.9 &    3.7 &  1.3  \nl
1690 - 1697 &  12.55 &  18:02:31.27 &  $-$30:04:36.2 &     0.7 &  1.5 &     2.2 &  1.5 &    2.3 &  2.1  \nl
1751 - 1752 &  14.06 &  18:02:09.44 &  $-$30:03:34.0 &  $-$6.6 &  1.1 &  $-$1.5 &  1.1 &    6.7 &  1.5  \nl
1773 - 1787 &  12.80 &  18:02:41.77 &  $-$30:03:04.0 &    11.1 &  1.8 &     5.8 &  1.8 &   12.6 &  2.6  \nl
1858 - 1859 &  14.02 &  18:02:39.34 &  $-$30:02:16.8 &     3.1 &  1.8 &     4.8 &  1.8 &    5.7 &  2.5  \nl
1873 - 1874 &  13.55 &  18:02:59.58 &  $-$30:02:48.0 &    10.9 &  1.1 & $-$25.2 &  1.1 &   27.5 &  1.6  \nl
1880 - 1887 &  14.56 &  18:03:04.10 &  $-$30:02:58.6 &  $-$6.6 &  2.2 & $-$27.8 &  2.1 &   28.6 &  3.0  \nl
1901 - 1910 &  13.54 &  18:02:21.52 &  $-$30:01:26.9 &  $-$8.2 &  1.0 &  $-$0.9 &  1.0 &    8.2 &  1.4  \nl
1986 - 1998 &  13.39 &  18:02:49.91 &  $-$30:01:06.4 &     1.9 &  1.4 &     1.9 &  1.4 &    2.7 &  1.9  \nl
2064 - 2065 &  13.58 &  18:02:04.65 &  $-$29:59:15.7 &     4.3 &  1.2 &    10.9 &  1.2 &   11.7 &  1.6  \nl
2071 - 2077 &  14.02 &  18:02:14.85 &  $-$29:58:52.9 &  $-$7.9 &  1.4 & $-$12.2 &  1.3 &   14.5 &  1.9  \nl
2157 - 2158 &  13.70 &  18:02:28.13 &  $-$29:58:01.5 &  $-$0.4 &  1.2 &  $-$8.6 &  1.2 &    8.6 &  1.7  \nl
2247 - 2257 &  13.41 &  18:02:33.95 &  $-$29:57:07.3 &     2.6 &  1.1 &     7.4 &  1.0 &    7.8 &  1.5  \nl
2255 - 2256 &  12.49 &  18:02:46.23 &  $-$29:57:27.2 &  $-$6.0 &  0.9 &     0.4 &  0.8 &    6.1 &  1.2  \nl
2391 - 2397 &  14.24 &  18:02:31.45 &  $-$29:55:26.6 &     9.3 &  1.8 &  $-$8.2 &  1.8 &   12.4 &  2.5  \nl
2525 - 2526 &  13.50 &  18:02:22.84 &  $-$29:54:06.9 &  $-$4.5 &  0.9 &  $-$2.7 &  0.9 &    5.3 &  1.3  \nl
2606 - 2607 &  13.95 &  18:02:34.76 &  $-$29:52:38.3 &  $-$3.1 &  1.2 & $-$13.4 &  1.2 &   13.7 &  1.7  \nl
2638 - 2639 &  14.12 &  18:03:04.01 &  $-$29:52:42.7 &    20.5 &  2.3 & $-$16.2 &  2.1 &   26.2 &  3.1  \nl
2677 - 2678 &  14.90 &  18:02:20.02 &  $-$29:51:37.3 &    13.0 &  1.4 &     4.9 &  1.4 &   13.9 &  2.0  \nl
2806 - 2816 &  13.51 &  18:02:37.58 &  $-$29:49:51.2 &  $-$7.6 &  1.0 &  $-$6.0 &  1.0 &    9.7 &  1.4  \nl
2988 - 2989 &  13.96 &  18:02:38.11 &  $-$29:48:42.7 &     8.8 &  1.5 &     3.7 &  1.4 &    9.5 &  2.1  \nl
3287 - 3300 &  15.09 &  18:02:23.29 &  $-$29:44:30.6 & $-$17.2 &  1.8 & $-$24.5 &  1.7 &   29.9 &  2.5  \nl
3378 - 3383 &  13.88 &  18:02:29.35 &  $-$29:44:11.4 &  $-$3.9 &  1.2 &  $-$1.9 &  1.2 &    4.3 &  1.7  \nl
3387 - 3388 &  13.12 &  18:02:34.95 &  $-$29:43:46.8 & $-$16.8 &  1.4 &  $-$8.8 &  1.4 &   18.9 &  2.0  \nl
3454 - 3455 &  13.28 &  18:02:41.66 &  $-$29:43:13.5 &     7.7 &  1.3 &     1.0 &  1.3 &    7.8 &  1.8  \nl
3501 - 3507 &  12.92 &  18:02:13.76 &  $-$29:42:04.6 &     2.4 &  1.4 &  $-$3.2 &  1.4 &    4.0 &  2.0  \nl
3667 - 3668 &  14.22 &  18:02:11.55 &  $-$29:40:42.5 &  $-$7.3 &  1.1 & $-$12.3 &  1.1 &   14.3 &  1.6  \nl
3670 - 3678 &  14.94 &  18:02:13.00 &  $-$29:40:11.9 & $-$16.5 &  1.9 &     5.0 &  1.8 &   17.2 &  2.6  \nl
3742 - 3750 &  13.93 &  18:02:11.69 &  $-$29:40:00.8 &  $-$5.6 &  1.4 &  $-$7.1 &  1.4 &    9.0 &  2.0  \nl
3743 - 3744 &  12.74 &  18:02:13.61 &  $-$29:39:18.2 &    62.6 &  1.2 & $-$23.6 &  1.2 &   66.9 &  1.7  \nl
3812 - 3813 &  11.80 &  18:02:19.07 &  $-$29:38:32.9 & $-$22.1 &  2.0 &    16.2 &  1.9 &   27.4 &  2.8  \nl
3855 - 3866 &  13.06 &  18:02:59.69 &  $-$29:38:30.8 &     5.9 &  1.3 &  $-$3.7 &  1.3 &    6.9 &  1.8  \nl
3893 - 3903 &  14.31 &  18:02:22.18 &  $-$29:37:38.0 &     4.2 &  1.4 &     8.3 &  1.4 &    9.3 &  2.0  \nl
3950 - 3951 &  14.16 &  18:02:07.51 &  $-$29:36:34.4 &  $-$1.9 &  1.1 &     6.6 &  1.1 &    6.8 &  1.6  \nl
3961 - 3970 &  13.75 &  18:02:31.02 &  $-$29:36:31.7 &     3.6 &  1.4 &     7.8 &  1.4 &    8.6 &  1.9  \nl
4003 - 4009 &  13.63 &  18:03:04.18 &  $-$29:37:18.4 & $-$11.8 &  1.9 &     5.2 &  1.8 &   12.8 &  2.6  \nl
4066 - 4067 &  13.96 &  18:02:53.60 &  $-$29:35:51.6 &  $-$1.2 &  1.4 & $-$24.7 &  1.4 &   24.7 &  1.9  \nl
4262 - 4263 &  13.80 &  18:02:23.86 &  $-$29:33:35.0 &    13.3 &  1.4 &     3.5 &  1.4 &   13.8 &  2.0  \nl
4304 - 4306 &  13.09 &  18:02:47.36 &  $-$29:33:02.5 &     6.7 &  1.4 &  $-$0.2 &  1.4 &    6.7 &  2.0  \nl
4479 - 4480 &  14.52 &  18:03:01.24 &  $-$29:31:53.3 &  $-$3.6 &  1.6 &     9.7 &  1.6 &   10.3 &  2.2  \nl

\enddata
\tablenotetext{\dagger}{$\mu_{\alpha\ast}=\mu_{\alpha\cos\delta}$}
\end{deluxetable}


\end{document}